\documentclass[preprint,amsmath,amssymb,11pt,english,aps,showkeys,showpacs]{revtex4}
\usepackage[english]{babel}
\usepackage{graphicx}
\usepackage{graphics}
\usepackage{amsmath}
\usepackage{dcolumn}
\usepackage{amssymb}
\usepackage{bm}
\usepackage{hyperref}

\begin{document}
\title{A massive scalar field under the effects of the Lorentz symmetry violation by a CPT-odd non-minimal coupling}

\author{R. L. L. Vit\'oria}
\email{ricardo.vitoria@ufes.br/ricardo.vitoria@pq.cnpq.br}
\affiliation{Departamento de F\'isica e Qu\'imica, Universidade Federal do Esp\'irito Santo, Av. Fernando Ferrari, 514, Goiabeiras, 29060-900, Vit\'oria, ES, Brazil.}

\author{H. Belich}
\email{belichjr@gmail.com}
\affiliation{Departamento de F\'isica e Qu\'imica, Universidade Federal do Esp\'irito Santo, Av. Fernando Ferrari, 514, Goiabeiras, 29060-900, Vit\'oria, ES, Brazil.}

\begin{abstract}
In this paper, based on the Standard Model Extended gauge sector, we made a non-minimal coupling in the Klein-Gordon equation which characterizes the Lorentz symmetry violation and, through this non-minimal CPT-odd coupling, we investigate the effects of possible scenarios of Lorentz symmetry violation by electrical and magnetic field configurations on a massive scalar field in this background, where, analytically, we determine solutions of bound states.
\end{abstract}


\maketitle

\section{Introduction}\label{sec1}

Recently, physical experiments have found supposed evidence that some particles detected in the Antarctic, despised as anomalies because they do not fit into theories, are real and must be taken into account \cite{sm, sm1}. In addition, in order to know better the atomic nucleus, researchers placed a muon in place of an orbital electron and showed that the proton radius is somewhat different from the theoretical predictions \cite{sm2}. Two decades ago, physicists have shown that one type of neutrino can transform into another, and for such a change to occur, it is necessary for the particle to have mass \cite{sm3} and that the fine structure constant, $\alpha=e^2/\hbar c$ \cite{sm4, sm5}. It is important to note that all these differences are in contrast to the predictions of the Standard Model (SM), which represents the best quantum field theory (QFT) that describes the subatomic world.

Due to the existence of conflicts between experimental/observational data and the theoretical predictions of the SM, the scientific community is induced to seek theories that best explain possible physical effects underlying the SM. In this context, the Lorentz symmetry violation (LSV) has been one of the alternative theories in an attempt to seek answers about the discrepancies between the experimental/observational results and the theoretical predictions of the SM. The maturation of the LSV culminated in a QFT that goes beyond the SM, which became known in the literature as Standard Model Extended (SME) \cite{sme, sme1}. The LSV has been studied in various branches of physics \cite{vsl, vsl1, vsl2, vsl3, vsl4, vsl5, vsl6, vsl7, vsl8, vsl9, vsl10, vsl11, vsl12, vsl13, vsl14, vsl15, vsl16, vsl17, vsl18, vsl19, vsl20, vsl21, vsl22, vsl23, vsl24, vsl25, vsl26, vsl27, vsl28, vsl29, vsl30, vsl31, vsl32, vsl33, vsl34, vsl35, vsl36, vsl37, vsl38, vsl39, vsl40, vsl41, vsl42, vsl43, vsl44, vsl45, vsl46, vsl47}. It is noteworthy that the LSV is also related to other theories, for example, the principle of generalized uncertainty \cite{new, new1} and rainbow gravity \cite{new2, new3}, where both imply a deformed Lorentz symmetry such that the energy-moment relations are modified in the plane spacetime by the corrections in the Plank scale. Recently, these two approaches, which modify the Klein-Gordon equation, have been investigated in black hole thermodynamics \cite{new4, new5}.

The LSV has been investigated in non-relativistic quantum mechanics, for example, on the influence of a Coulomb-like potential induced by the Lorentz symmetry breaking effects on the harmonic oscillator \cite{bb}, on geometric phases for a Dirac neutral particle \cite{bb1, bb2}, in a Rashba coupling induced by LSV effects \cite{bb3}, in quantum holonomies, on a Dirac neutral particle inside a two-dimensional quantum ring \cite{bb4}, in a Landau-type quantization \cite{bb5}, in a spin-orbit coupling for a neutral particle \cite{bb6}, in a Rashba-type coupling induced by Lorentz-violating effects on a Landau system for a neutral particle \cite{bb7} and on an Aharonov-Bohm effect induced by the LSV \cite{bf}. In relativistic quantum mechanics, the LSV has been studied in a  relativistic Anandan quantum phase \cite{bb8}, in a relativistic Landau-Aharonov-Casher quantization \cite{bb9}, in a relativistic Landau-He-McKellar-Wilkens quantization and relativistic bound states solutions for a Coulomb-like potential induced by the Lorentz symmetry breaking effects \cite{bb10}, on geometric quantum phases from Lorentz symmetry breaking effects in the cosmic string spacetime \cite{bb1}, on relativistic EPR correlations \cite{bb12}, on the relativistic Anandan quantum phase and the Aharonov-Casher effect under Lorentz symmetry breaking effects in the cosmic string spacetime \cite{bb13} and in a relativistic quantum scattering yielded by Lorentz symmetry breaking effects \cite{bb14}. All of these examples are investigations of a spin-$1/2$ fermionic field. In addition, these studies were possible through non-minimal couplings in the Dirac equation.

Non-minimum couplings, based on the SME gauge sector, which carry information that the background is characterized by the LSV are classified in the literature as CPT-even and CPT-odd. The first conserves the CPT symmetry, while the second violates the CPT symmetry \cite{book}. Recently, the calibre sector CPT-even coupling has been investigated on a scalar field in solutions of bound states, for example, on a relativistic scalar particle subject to a Coulomb-type potential given by Lorentz symmetry breaking effects \cite{bb15}, on the harmonic-type and linear-type confinement of a relativistic scalar particle yielded by Lorentz symmetry breaking effects \cite{bb16}, in a relativistic scalar particle subject to a confining potential and Lorentz symmetry breaking effects in the cosmic string spacetime \cite{bb17}, on the effects of the Lorentz symmetry violation yielded by a tensor field on the interaction of a scalar particle and a Coulomb-type field \cite{me, me1} and on the Klein-Gordon oscillator \cite{me2, me3}. However, a point that has not yet been raised in the literature, in the context of relativistic quantum mechanics, is the CPT-odd non-minimal coupling, based on the SME gauge sector, in the Klein-Gordon equation. Therefore, following Refs. \cite{hb, hb1}, we insert a background vector field, which governs the LSV, into the Klein-Gordon equation through a non-minimal coupling. Next, we consider distributions of electric and magnetic fields to which they characterize possible scenarios of LSV where it is possible to obtain solutions of bound states for such backgrounds. In addition, we consider one of these backgrounds, the most general, and we analyze a massive scalar field with position-dependent mass.

The structure of this paper is as follows: in the Sec. (\ref{sec2}), we rewrite the Klein-Gordon equation under the perspective of a non-minimal coupling that carries the information that the LSV is present by the a background vector field. Next, we analyze the first LSV scenario possible through an electric field configuration which induces a Coulomb-type potential and determine the relativistic energy levels of the system. Continuing, we investigated the second LSV possible scenario where there is a uniform magnetic field, where we determine analytically the energy spectrum of the system; in the Sec. (\ref{sec3}), still in the second background, we insert a linear central potential in the Klein-Gordon equation by modifying the mass term and show that the relativistic energy levels of the system are drastically modified; in the Sec. (\ref{sec4}), we present our conclusions.

\section{CPT-old non-minimal coupling in the Klein-Gordon equation}\label{sec2}

The description of a massive scalar field is given by the Klein-Gordon equation
\begin{eqnarray}\label{eq01}
\Box\phi-m^2\phi=0,
\end{eqnarray}
where $\Box=\partial_{\mu}\partial^{\mu}$, with $\mu=0,i=1,2,3$, and $m$ is rest mass of the scalar field $\phi$. By considering the Minkowski spacetime with cylindrical symmetry described by the metric $(c=\hbar=1)$
\begin{eqnarray}\label{eq02}
ds^2=-dt^2+d\rho^2+\rho^2d\varphi^2+dz^2,
\end{eqnarray}
where $\rho=\sqrt{x^2+y^2}$, the Klein-Gordon equation is given as follows
\begin{eqnarray}\label{eq03}
-\frac{\partial^2\phi}{\partial t^2}+\frac{\partial^2\phi}{\partial\rho^2}+\frac{1}{\rho}\frac{\partial\phi}{\partial\rho}+\frac{1}{\rho^2}\frac{\partial^2\phi}{\partial\varphi^2}
+\frac{\partial^2\phi}{\partial z^2}-m^2\phi=0,
\end{eqnarray}
which represents the relativistic quantum dynamics of a free scalar field in the Minkowski spacetime, that is, in an isotropic medium. On the other hand, we can leave an isotropic medium to an anisotropic medium through a non-minimal coupling in the Klein-Gordon equation which is characterized by the presence of background vector and tensor fields that governs the LSV \cite{sme, sme1, book}. Then, based on the Refs. \cite{hb, hb1}, let us consider the CPT-old non-minimum coupling $\partial_{\mu}-ig\tilde{F}_{\mu\alpha}v^{\alpha}$, where $g\ll1$ is a coupling constant, $\tilde{F}_{\mu\nu}=\frac{1}{2}\varepsilon_{\mu\nu\alpha\beta}F^{\alpha\beta}$ is dual electromagnetic tensor, $F_{\mu\nu}=\partial_{\mu}A_{\nu}-\partial_{\nu}A_{\mu}$ is the Maxwell tensor and $v^{\alpha}$ is the background constant vector field that governs the LSV. In the way, the Eq. (\ref{eq01}) becomes
\begin{eqnarray}\label{eq04}
(\partial_{\mu}-ig\tilde{F}_{\mu\alpha}v^{\alpha})(\partial^{\mu}-ig\tilde{F}^{\mu\beta}v_{\beta})\phi-m^2\phi=0,
\end{eqnarray}
or
\begin{eqnarray}\label{eq05}
\Box\phi-2ig(\partial_{\mu}\phi)\tilde{F}^{\mu\alpha}v_{\alpha}-ig(\partial_{\mu}\tilde{F}^{\mu\alpha})v_{\alpha}
-g^2\tilde{F}_{\mu\alpha}v^{\alpha}\tilde{F}^{\mu\alpha}v_{\alpha}-m^2\phi=0.
\end{eqnarray}

It is worth remembering that $\partial_{\mu}\tilde{F}^{\mu\alpha}=0$ gives the homogenous Maxwell equations. Moreover, since $g^2v^{\alpha}v_{\alpha}\ll1$, we neglect the penultimate term of the Eq. (\ref{eq05}) such that we obtain
\begin{eqnarray}\label{eq06}
\Box\phi-2ig(\partial_{\mu}\phi)\tilde{F}^{\mu\alpha}v_{\alpha}-m^2\phi=0.
\end{eqnarray}

Hence, in a spacetime described by the metric given in the Eq. (\ref{eq02}), the Eq. (\ref{eq06}) is given in the form
\begin{eqnarray}\label{eq07}
-\frac{\partial^2\phi}{\partial t^2}+\frac{\partial^2\phi}{\partial\rho^2}+\frac{1}{\rho}\frac{\partial\phi}{\partial\rho}+\frac{1}{\rho^2}\frac{\partial^2\phi}{\partial\varphi^2}
+\frac{\partial^2\phi}{\partial z^2}+2ig\vec{v}.\vec{B}\partial^0\phi+2igv^0\vec{B}.\nabla\phi-2ig(\vec{v}\times\vec{E}).\nabla\phi-m^2\phi=0.
\end{eqnarray}

The Eq. (\ref{eq07}) represents the Klein-Gordon equation in a spacetime of cylindrical symmetry affected by the LSV, which is governed by the presence of a background vector field. Note that the Eq. (\ref{eq07}) supports several LSV scenarios through possible electric and magnetic field configurations. We will investigate two particular cases from now on.

\subsection{First background}\label{sec2a}

Let us consider a background of the LSV determined by the field configuration:
\begin{eqnarray}\label{eq08}
v_{\alpha}=(0,0,v_{\varphi},0); \ \ \ \vec{E}=\frac{\lambda}{\rho}\hat{\rho}; \ \ \ \vec{B}=0,
\end{eqnarray}
where $v_{\varphi}=\text{const.}$ and $\lambda$ is a constant associated with a linear distribution of electric charges on the $z$-axis. Note that, with this field configuration, we have $\vec{v}\times\vec{E}=-\frac{v_{\varphi}\lambda}{\rho}\hat{z}$. Then, for this possible scenario, the Eq. (\ref{eq07}) becomes
\begin{eqnarray}\label{eq09}
-\frac{\partial^2\phi}{\partial t^2}+\frac{\partial^2\phi}{\partial\rho^2}+\frac{1}{\rho}\frac{\partial\phi}{\partial\rho}+\frac{1}{\rho^2}\frac{\partial^2\phi}{\partial\varphi^2}
+\frac{\partial^2\phi}{\partial z^2}+\frac{2igv_{\varphi}\lambda}{\rho}\frac{\partial\phi}{\partial z}-m^2\phi=0.
\end{eqnarray}

The solution to the Eq. (\ref{eq09}) is given in form
\begin{eqnarray}\label{eq10}
\phi(\rho,\varphi,z,t)=R(\rho)e^{-i(\mathcal{E}t-l\varphi-kz)},
\end{eqnarray}
where $l=0,\pm1,\pm2,\ldots$ and $-\infty<k<\infty$ are the eigenvalues of the angular momentum $\hat{L}_z=-i\partial_{\varphi}$ and linear $\hat{p}_{z}=-i\partial_{z}$ operators, respectively. Then, by substituting the Eq. (\ref{eq10}) into the Eq. (\ref{eq09}), we obtain
\begin{eqnarray}\label{eq11}
\frac{d^2R}{d\rho^2}+\frac{1}{\rho}\frac{dR}{d\rho}-\frac{l^2}{\rho^2}R-\frac{2gkv_{\varphi}\lambda}{\rho}R+(\mathcal{E}^2-m^2-k^2)R=0.
\end{eqnarray}

Now, let us assume the case $\lambda=-|\lambda|$. In the way, the Eq. (\ref{eq11}) is rewrite in the form
\begin{eqnarray}\label{eq12}
\frac{d^2R}{d\rho^2}+\frac{1}{\rho}\frac{dR}{d\rho}-\frac{l^2}{\rho^2}R+\frac{2gkv_{\varphi}|\lambda|}{\rho}R-a^2R=0,
\end{eqnarray}
where we define the parameter
\begin{eqnarray}\label{eq13}
a^2=m^2+k^2-\mathcal{E}^2.
\end{eqnarray}

By making the change of variable $\xi=2a\rho$ into Eq. (\ref{eq12}), we have
\begin{eqnarray}\label{eq14}
\frac{d^2R}{d\xi^2}+\frac{1}{\xi}\frac{dR}{d\xi}-\frac{l^2}{\xi^2}R+\frac{b}{\xi}R-\frac{1}{4}R=0,
\end{eqnarray}
with
\begin{eqnarray}\label{eq15}
b=\frac{gkv_{\varphi}|\lambda|}{a}.
\end{eqnarray}

The radial wave function $R(\xi)$ must be analytic at the origin and at the infinity. Then, by analyzing the asymptotic behavior of the Eq. (\ref{eq14}) at $\xi\rightarrow0$ and $\xi\rightarrow\infty$, we obtain the following solution in terms of a function $f(\xi)$:
\begin{eqnarray}\label{eq16}
R(\xi)=\xi^{l}e^{-\frac{\xi}{2}}f(\xi).
\end{eqnarray}

By substituting the Eq. (\ref{eq16}) into the Eq. (\ref{eq14}), we have
\begin{eqnarray}\label{eq17}
\xi\frac{d^2f}{d\xi^2}+(2|l|+1-\xi)\frac{df}{d\xi}+\left(b-|l|-\frac{1}{2}\right)f=0.
\end{eqnarray}

The Eq. (\ref{eq17}) is the well known confluent hypergeometric equation \cite{arf, abr} and $f(\xi)$ is the confluent hypergeometric power series: $f(\xi)= \ _{1}F_{1}\left(|l|+\frac{1}{2}-b,2|l|+1;\xi\right)$. It is well known that the confluent hypergeometric power series becomes a polynomial of degree $n=0,1,2,\ldots$ when we have the condition $|l|+\frac{1}{2}-b=-n$. Then, with this condition, we obtain
\begin{eqnarray}\label{eq18}
\mathcal{E}_{k,l,n}=\pm\sqrt{m^2+\left[1-\frac{g^2v_{\varphi}^2\lambda^2}{\left(n+|l|+\frac{1}{2}\right)^2}\right]k^2},
\end{eqnarray}
which represents the relativistic energy levels of a massive scalar field in a LSV background governed for a constant vector field. This energy spectrum arises through a Coulomb-type central potential induced by the LSV, by the electric and magnetic field configuration given in the Eq. (\ref{eq08}). Hence, the energy spectrum is influenced by the LSV through parameters $g$, $v_{\varphi}$ and $\lambda$. We can note that for the particular case where $k=0$, that is, the scale field is in the $xy$-plane, we obtain the rest energy of the scalar field, in contrast to Ref. \cite{bb16}, where the scalar field, which is subjected to the CPT-even non-minimal coupling of the SME gauge sector, remains confined in the $xy$-plane. This effect may be associated with the non-minimal CPT-odd coupling. We can also notice that, for $g\rightarrow0$, we recover the energy corresponding to a free scalar field in the Minkowski spacetime.

\subsection{Second background}\label{sec2b}

In this section, we establish another LSV scenario possible. This LSV scenario possible is determined by a field configuration defined as
\begin{eqnarray}\label{eq19}
v_{\alpha}=(0,0,v_{\varphi},v_{z}); \ \ \ \vec{E}=\frac{\lambda}{\rho}\hat{\rho}; \ \ \ \vec{B}=B_0\hat{z},
\end{eqnarray}
where $B_{0}$ is a constant. Note that $\vec{v}\times\vec{E}=\frac{\lambda}{\rho}(v_z\hat{\varphi}-v_{\varphi}\hat{z})$. In the way, the Eq. (\ref{eq07}) becomes
\begin{eqnarray}\label{eq20}
-\frac{\partial^2\phi}{\partial t^2}+\frac{\partial^2\phi}{\partial\rho^2}+\frac{1}{\rho}\frac{\partial\phi}{\partial\rho}+\frac{1}{\rho^2}\frac{\partial^2\phi}{\partial\varphi^2}
+\frac{\partial^2\phi}{\partial z^2}+2igB_0v_{z}\frac{\partial\phi}{\partial t}-\frac{2igv_{z}\lambda}{\rho^2}\frac{\partial\phi}{\partial\varphi}+\frac{2igv_{\varphi}\lambda}{\rho}\frac{\partial\phi}{\partial z}-m^2\phi=0.
\end{eqnarray}

By following the steps from the Eq. (\ref{eq10}) to the Eq. (\ref{eq12}), we have
\begin{eqnarray}\label{eq21}
\frac{d^2R}{d\rho^2}+\frac{1}{\rho}\frac{dR}{d\rho}-\frac{\gamma^2}{\rho^2}R+\frac{2gkv_{\varphi}|\lambda|}{\rho}R-c^2R=0,
\end{eqnarray}
with the parameters
\begin{eqnarray}\label{eq22}
c=m^2+k^2-\mathcal{E}^2-2gB_0v_z\mathcal{E}; \ \ \ \gamma^2=l^2-2glv_z|\lambda|.
\end{eqnarray}

By making the change of variable $\varrho=2c\rho$ into Eq. (13), we obtain
\begin{eqnarray}\label{eq23}
\frac{d^2R}{d\varrho^2}+\frac{1}{\varrho}\frac{dR}{d\varrho}-\frac{\gamma^2}{\varrho^2}R+\frac{d}{\varrho}R-\frac{1}{4}R=0,
\end{eqnarray}
where we define the new parameter
\begin{eqnarray}\label{eq24}
d=\frac{gkv_{\varphi}|\lambda|}{c}.
\end{eqnarray}

We can note that the Eq. (\ref{eq23}) is analogous to the Eq. (\ref{eq14}). Then, by analyzing the asymptotic behavior of the Eq. (\ref{eq23}) at $\varrho\rightarrow0$ and at $\varrho\rightarrow\infty$, we obtain a general solution in terms of a confluent hypergeometric function: $R(\varrho)=\varrho^{|\gamma|}e^{-\frac{1}{2}\varrho} \ _{1}F_{1}\left(|\gamma|+\frac{1}{2}-d,2|\gamma|+1;\varrho\right)$. As we saw in the previous case, in order to obtain relativistic bound state solutions, we must impose the condition $|\gamma|+\frac{1}{2}-d=-n$, with $n=0,1,2,\ldots$, which gives us
\begin{eqnarray}\label{eq25}
\mathcal{E}_{k,l,n}=-gB_0v_{z}\pm\sqrt{g^2B_0^2v_z^2+m^2+\left[1-\frac{g^2v_{\varphi}^2\lambda^2}{\left(n+\sqrt{l^2-2glv_z|\lambda|}+\frac{1}{2}\right)^2}\right]k^2}.
\end{eqnarray}

The Eq. (\ref{eq25}) represents the energy spectrum of massive scalar field in a LSV background governed for a constant vector field. We can note that the new LSV background determined by the configuration given in the Eq. (\ref{eq19}) modifies relativistic energy levels, compared to the Eq. (\ref{eq18}). This modification can be seen by the presence of parameters $v_z$ and $B_0$ in the Eq. (\ref{eq25}). Again, we can also note that, for the particular case where $k=0$, the massive scalar field is not confined in the $xy$-plane, as in the Ref. \cite{bb16}. In addition, in this particular case, in contrast to the Eq. (\ref{eq18}), the massive scalar field gains a relativistic energy term determined by the parameters associated with the LSV background on the rest energy, that is, $\mathcal{E}=-\epsilon_{LSV}\pm\sqrt{\epsilon_{LSV}^2+m^2}$, where $\epsilon_{LSV}=gB_0v_z$. By making $g\rightarrow0$ in the Eq. (\ref{eq25}), we recover the energy corresponding to a free massive scalar field in the Minkowski space.

\section{Effects of a linear central potential}\label{sec3}

The Ref. \cite{greiner} presents a procedure to insert central potentials in the Klein-Gordon equation different from the standard procedure that is through the minimum coupling $\hat{p}_{\mu}\rightarrow\hat{p}_{\mu}-qA_{\mu}$, where $q$ is the electric charge of the scalar field. This other procedure of inserting central potentials is done by modifying the mass term of the Klein-Gordon equation by the transformation $m\rightarrow m+S(\vec{r})$, where $S(\vec{r})=S(\rho)$ is the scalar central potential. Then, through this procedure, let us insert a central linear potential in the Klein-Gordon equation in the form:
\begin{eqnarray}\label{eq26}
m\rightarrow m+\eta\rho,
\end{eqnarray}
where $\eta$ is a constant that characterizes the linear central potential. In the context of non-relativistic quantum mechanical, the linear central potential has been studied on the effects of a screw dislocation on the harmonic oscillator \cite{pl}, on an atom with a magnetic quadrupole moment \cite{pl1}, in atomic and molecular physics \cite{pl2, pl3, pl4, pl5, pl6, pl7} and on a quantum particle subject to the uniform force field \cite{landau, pl8}; in the context of relativistic quantum mechanics, the liner central potential has been studied in atomic physics \cite{pl9}, in the Klein-Gordon equation in the presence of a dyon and magnetic flux in the spacetime of gravitational defects \cite{pl10}, in the relativistic quantum dynamics of a charged particle in cosmic
string spacetime in the presence of magnetic field \cite{eug}, on Klein-Gordon oscillator \cite{me4, me5}, on a scalar field in the spacetime with torsion \cite{me6, me7, me8}, on a Majorana fermion \cite{pl11}, on a scalar field in the rotating cosmic string spacetime \cite{pl12, pl13} and on a scalar particle in a G\"odel-type spacetime \cite{pl14, me9}.

By substituting the Eq. (\ref{eq26}) into the Eq. (\ref{eq07}) and by considering the background given in the subsection (\ref{sec2b}), we have the radial wave equation
\begin{eqnarray}\label{eq27}
\frac{d^2R}{d\rho^2}+\frac{1}{\rho}\frac{dR}{d\rho}-\frac{\gamma^2}{\rho^2}R-\frac{2gkv_{\varphi}\lambda}{\rho}R-2m\eta\rho R-\eta^2\rho^2R+\bar{c}^2R=0,
\end{eqnarray}
where
\begin{eqnarray}\label{eq28}
\bar{c}^2=\mathcal{E}^2+2gB_0v_z\mathcal{E}-m^2-k^2.
\end{eqnarray}

From now on, let us consider the change of variable $\zeta=\sqrt{\eta}\rho$, such that the Eq. (\ref{eq27}) becomes
\begin{eqnarray}\label{eq29}
\frac{d^2R}{d\zeta^2}+\frac{1}{\zeta}\frac{dR}{d\zeta}-\frac{\gamma^2}{\zeta^2}R-\frac{\theta}{\zeta}R-\kappa\zeta R-\zeta^2R+\frac{\bar{c}^2}{\eta}R=0,
\end{eqnarray}
where we define the new parameters
\begin{eqnarray}\label{eq30}
\theta=\frac{2gkv_{\varphi}\lambda}{\sqrt{\eta}}; \ \ \ \kappa=\frac{2m}{\sqrt{\eta}}.
\end{eqnarray}

Again, we are interested in analytical solutions for $\zeta\rightarrow0$ and $\zeta\rightarrow\infty$. In this sense, by analyzing the asymptotic behavior of the Eq. (\ref{eq29}) for $\zeta\rightarrow0$ and $\zeta\rightarrow\infty$, we can write a general solution in terms of a unknown function $g(\zeta)$ in the form \cite{eug}:
\begin{eqnarray}\label{eq31}
R(\zeta)=\zeta^{|\gamma|}e^{-\frac{1}{2}\zeta(\zeta+\kappa)}g(\zeta).
\end{eqnarray}

By substituting the Eq. (\ref{eq31}) into the Eq. (\ref{eq29}), we obtain
\begin{eqnarray}\label{eq32}
\frac{d^2g}{d\zeta^2}+\left[\frac{(2|\gamma|+1)}{\zeta}-\kappa-2\zeta\right]\frac{dg}{d\zeta}+\left[\sigma-\frac{\tau}{\zeta}\right]g=0,
\end{eqnarray}
where
\begin{eqnarray}\label{eq33}
\sigma=\frac{\bar{c}^2}{\eta}+\frac{\kappa^2}{4}-2(1+|\gamma|); \ \ \ \tau=\theta+\frac{\kappa}{2}(2|\gamma|+1).
\end{eqnarray}

The Eq. (\ref{eq32}) is the biconfluent Heun equation \cite{eug, heun} and $g(\zeta)$ is the biconfluent Heun function: $g(\zeta)=H_{b}\left(2|\gamma|,\kappa,\frac{\bar{c}^2}{\eta}+\frac{\kappa^2}{4},2\theta;\zeta\right)$. The Eq. (\ref{eq32}) has two singular points: $\zeta=0$ and $\zeta\rightarrow\infty$. $\zeta=0$ represents the regular singular point, while $\zeta\rightarrow\infty$ represents the irregular singular point \cite{eug}. In this case, the Eq. (\ref{eq32}) has at least one solution around the origin given by the power series \cite{arf, eug}
\begin{eqnarray}\label{eq34}
g(\zeta)=\sum_{j=0}^{\infty}s_{j}\zeta^{j}.
\end{eqnarray}

By substituting the Eq. (\ref{eq34}) into the Eq. (\ref{eq32}), we obtain recurrence relation
\begin{eqnarray}\label{eq35}
s_{j+2}=\frac{[\tau+\kappa(j+1)]s_{j+1}-(\sigma-2j)s_j}{(j+2)(j+2+2|\gamma|)},
\end{eqnarray}
with the coefficients
\begin{eqnarray}\label{eq36}
s_{1}=\frac{\tau}{(1+2|\gamma|)}s_0; \ \ \ s_{2}=\frac{s_0}{4(1+|\gamma|)}\left[\frac{(\tau+\kappa)}{(1+2|\gamma|)}-\sigma\right].
\end{eqnarray}

By following Refs. \cite{pl10, eug, me9, heun1}, it is possible to obtain solutions of bound states and, consequently, a polynomial of degree $\bar{n}=1,2,3,\ldots$ of the Eq. (\ref{eq34}) imposing the following conditions:
\begin{eqnarray}\label{eq37}
\sigma=2\bar{n}; \ \ \ s_{\bar{n}+1}=0.
\end{eqnarray}

Let us consider the lowest energy state of the system, that is, the radial mode $\bar{n}=1$ in the condition $s_{\bar{n}+1}=0$ such that we obtain $s_2=0$. For this equality to give us a physical meaning, an adjustment parameter is needed, not only for the radial mode $\bar{n}=1$, but for any value of $\bar{n}$. Then, we choose the parameter that characterizes the linear central potential $\eta$ as the adjustment parameter, which, from the $s_2=0$, gives the following expression:
\begin{eqnarray}\label{eq38}
\eta_{k,l,1}=\frac{m^2}{2}(2|\gamma|+3)+4gkmv_{\varphi}\lambda\frac{(|\gamma|+1)}{(2|\gamma|+1)}+\frac{2g^2k^2v_{\varphi}^2\lambda^2}{(2|\gamma|+1)}.
\end{eqnarray}

The Eq. (\ref{eq38}) represents the allowed values of the parameter associated to the linear central potential for the radial mode $\bar{n}=1$. We can observe that the parameter $\eta$ is influenced by the LSV through the presence of the parameters $g$, $\lambda$, $v_{\varphi}$ and $v_{z}$ in the Eq. (\ref{eq38}). In addition, the specific values of the parameter $\eta$ depend on the quantum numbers $\{k,l,\bar{n}\}$ and for this reason, we have labelled $\eta=\eta_{k,l,\bar{n}}$ in the Eq. (\ref{eq38}). We can note that, for the particular case $k=0$, the Eq. (\ref{eq38}) becomes $\eta_{k,l,1}=\frac{m^2}{2}(2\sqrt{l^2-2glv_{z}\lambda}+3)$, which represents the allowed values of the parameter $\eta$ for radial mode $\bar{n}=1$, that is, the allowed values of $\eta$ for the radial mode $\bar{n}=1$ continue to be influenced by the LSV through the parameters $g$ and $v_z$. We can also note that, for $g\rightarrow0$, we recover the expression given in the Ref. \cite{me6} without topological defect.

For our analysis to become complete, it is necessary to analyze the condition $\sigma=2\bar{n}$ for radial mode $\bar{n}=1$. Then, the condition $\sigma=2\bar{n}$ becomes $\sigma=2$, which gives us the expression
\begin{eqnarray}\label{eq39}
\mathcal{E}_{k,l,1}=-gB_0v_z\pm\sqrt{g^2B_0^2v_z^2+k^2+2\eta_{k,l,1}(2+|\gamma|)}.
\end{eqnarray}
Then, by substituting the Eq. (\ref{eq38}) into the Eq. (\ref{eq39}), we obtain
\begin{eqnarray}\label{eq40}
\mathcal{E}_{k,l,1}=&-&gB_0v_z \nonumber \\
&\pm&\sqrt{g^2B_0^2v_z^2+k^2+\left[m^2(2|\gamma|+3)+8gkmv_{\varphi}\lambda\frac{(|\gamma|+1)}{(2|\gamma|+1)}
+\frac{4g^2k^2v_{\varphi}^2\lambda^2}{(2|\gamma|+1)}\right](2+|\gamma|)}.
\end{eqnarray}

The Eq. (\ref{eq40}) (or the Eq. (\ref{eq39})) represents the allowed values of relativistic energy of a massive scalar field to the lowest energy state subject to a linear central potential in a LSV possible scenario governed by a background vector field. We can note that the allowed values of relativistic energy for the state of lower energy of the system are influenced by the LSV through the parameters associated with the background considered $(g,B_0,v_{\varphi},v_z,\lambda)$. In contrast to the Eqs. (\ref{eq18}) and (\ref{eq25}), it is not possible to obtain a closed equation representing the relativistic energy spectrum, but rather separate expressions representing the allowed values of relativistic energy for the radial modes. This effect is due to the presence of the linear central potential in the system, which has its parameter dependent on the quantum numbers. We can also note that, for the particular case $k=0$, in contrast to the Eqs. (\ref{eq18}) and (\ref{eq25}), the massive scalar field remains confined in the plane and influenced by the LSV through the parameters $g$, $B_0$ and $v_z$. Again, this effect is due to the presence of the linear central potential. By making $g\rightarrow0$, we recover the allowed values of relativistic energy for radial mode $\bar{n}=1$ obtained in the Ref. \cite{me6} without torsion.

\section{Conclusion}\label{sec4}

We have investigated a massive scalar field subject to the effects of the LSV through a non-minimal coupling based on the SME gauge sector. We have shown that through this minimum coupling the Klein-Gordn equation is modified and, consequently, provides several LSV backgrounds to be investigated. In this sense, we investigated two backgrounds, where we can observe that in both a Coulomb-type central potential is induced by the LSV and, therefore, we determine the relativistic energy spectra, which are influenced by the LSV.

In addition, we introduce a linear central potential in the Klein-Gordon equation by modifying the mass term, where we analyze the effects of this central potential on the second background. We can note that the presence of this potential modifies the relativistic energy levels of the system. This modification is characterized by the impossibility of determining a closed expression representing the relativistic energy spectrum of the system. Given this, we have shown that it is possible to determine allowed values of relativistic energy for each radial mode of the system separately, where these allowed values are influenced by the LSV. As an example, we determine the allowed values of relativistic energy for the lower energy state, that is, radial mode $\bar{n}=1$, instead of $n=0$. Another interesting quantum effect is that the parameter associated with the linear central potential is determined by the LSV and on the quantum numbers of the system.

\acknowledgments{The authors would like to thank CNPq (Conselho Nacional de Desenvolvimento Cient\'ifico e Tecnol\'ogico - Brazil). Ricardo L. L. Vit\'oria was supported by the CNPq project No. 150538/2018-9.}

\end{document}